\documentclass[]{aa}

\usepackage{txfonts}
\usepackage{graphicx}
\usepackage{aalongtable}

\usepackage{natbib}

\bibpunct{(}{)}{;}{a}{}{,}

\begin{document}

\title{Velocity-metallicity correlation for high-$z$ DLA galaxies:\\
Evidence for
a mass-metallicity relation ?\thanks{Based on data gathered at the
European Southern Observatory (ESO) using the Ultraviolet and Visible Echelle
Spectrograph (UVES) installed at the Very Large Telescope (VLT), Unit-2,
Kueyen, on Cerro Paranal in Chile.}}

\titlerunning{Velocity-metallicity correlation for high-$z$ DLA galaxies}

\author{C. Ledoux\inst{1},
        P. Petitjean\inst{2,3},
        J.~P.~U. Fynbo\inst{4},
        P. M\o ller\inst{5},
        and R. Srianand\inst{6}}

%\offprints{C. Ledoux, \email{cledoux@eso.org}}

\institute{
   European Southern Observatory, Alonso de C\'ordova 3107, Casilla 19001, Vitacura, Santiago 19, Chile\\
   \email{cledoux@eso.org}
\and
   Institut d'Astrophysique de Paris -- UMR 7095 CNRS \& Universit\'e Pierre et Marie Curie, 98bis Boulevard Arago, 75014 Paris, France\\
   \email{petitjean@iap.fr}
\and
   LERMA, Observatoire de Paris, 61 Avenue de l'Observatoire, 75014 Paris, France
\and
   Dark Cosmology Centre, Niels Bohr Institute, University of Copenhagen, Juliane Maries Vej 30, 2100 Copenhagen {\O}, Denmark\\
   \email{jfynbo@astro.ku.dk}
\and
   European Southern Observatory, Karl-Schwarzschild-Stra\ss e 2, 85748 Garching bei M\"unchen, Germany\\
   \email{pmoller@eso.org}
\and
   Inter-University Centre for Astronomy and Astrophysics, Post Bag 4, Ganesh Khind, Pune 411 007, India\\
   \email{anand@iucaa.ernet.in}
}

\date{Received date / Accepted date}

\abstract{
We used our database of VLT-UVES quasar spectra to build up a sample of
70 Damped Lyman-$\alpha$ (DLA) or strong sub-DLA systems with total
neutral hydrogen column densities of $\log N($H\,{\sc i}$)\ga 20$ and redshifts
in the range $1.7<z_{\rm abs}<4.3$. For each of the systems, we measured in an
homogeneous manner the metallicities relative to Solar, [X/H] (with
${\rm X}={\rm Zn}$, or S or Si), and the velocity widths of low-ionization line
profiles, $\Delta V$. We provide for the first time evidence for a
correlation between DLA metallicity and line profile velocity width, which
is detected at the $6.1\sigma$ significance level. This confirms the trend
previously observed in a much smaller sample by \citet{1998ApJ...494L..15W}.
The best-fit linear relation is
$[{\rm X}/{\rm H}]=1.55(\pm 0.12)\log\Delta V -4.33(\pm 0.23)$ with $\Delta V$
expressed in km s$^{-1}$. The slope of the DLA velocity-metallicity relation is
the same within uncertainties between the higher ($z_{\rm abs}>2.43$) and the
lower ($z_{\rm abs}\le 2.43$) redshift halves of our sample. However, the two
populations of systems are statistically different. There is a strong
redshift evolution in the sense that the mean metallicity {\sl and}
mean velocity width increase with decreasing redshift. We argue that
the existence of a DLA velocity-metallicity correlation, over more than a
factor of 100 spread in metallicity, is probably the consequence of an
underlying mass-metallicity relation for the galaxies responsible for DLA
absorption lines. Assuming a simple linear scaling of the galaxy
luminosity with the mass of the dark-matter halo, we find that the slope of the
DLA velocity-metallicity relation is consistent with that of the
luminosity-metallicity relation derived for local galaxies. If the galaxy
dynamical mass is indeed the dominant factor setting up the observed DLA
velocity-metallicity correlation, then the DLA systems exhibiting the lowest
metallicities among the DLA population should, on average, be associated
with galaxies of lower masses (e.g., gas-rich dwarf galaxies). In turn, these
galaxies should have the lowest luminosities among the DLA galaxy
population. This could explain the difficulties of detecting high-redshift
DLA galaxies in emission.
}

\maketitle

\section{Introduction}

Over the past decade, significant progress in our understanding of early galaxy
evolution has been made with large samples of high-redshift galaxies drawn
from deep multi-band imaging \citep[][and
references therein]{2003ApJ...592..728S}. However, even before the
first surveys for Lyman-Break Galaxies (LBGs) had begun, samples of DLA
absorbers observed on the lines-of-sight to distant quasars had been
constructed \citep{1986ApJS...61..249W,1995ApJ...454..698W}. These
absorbers were thought at the time to be the best carriers of information on
the population of high-redshift galaxies, but, despite many attempts
to identify the galaxies responsible for DLA absorption lines [hereafter
called DLA galaxies], only very few could be detected in emission \citep[see,
e.g.,][]{1993A&A...270...43M,1996Natur.382..234D,1999MNRAS.303..711L,1999MNRAS.305..849F,2002ApJ...574...51M,2004A&A...422L..33M,2004A&A...417..487C,2005MNRAS.358..985W}.
However, there is little doubt that DLA systems arise from the densest
regions of the Universe and are closely associated with galaxies. It is
therefore crucial to establish the connection between the absorption-selected
DLA systems and emission-selected galaxies. In addition, the detailed
information that becomes available only through the combination of morphology,
colour and luminosity, with QSO absorption-line spectroscopy, makes these
galaxy/absorber associations unique laboratories to study the
physical processes at work during galaxy
formation \citep[see][]{1999ApJ...522..604P}.

Progress in this field has been slow. Firstly, a huge amount of work is
needed to derive important parameters in DLA systems such as gas kinematics,
metallicity, or dust and molecular fractions
\citep[see, e.g.,][]{1997ApJ...486..665P,1997ApJ...487...73P,1998ApJ...507..113P,1999ApJS..121..369P,2001ApJ...552...99P,1998A&A...337...51L,2003MNRAS.346..209L}.
Secondly, as mentioned above, the known high-redshift DLA systems have
proved to be very difficult to detect in emission. This has caused
some confusion and for a while suggestions were put forward that DLA
absorbers may not be related to high-redshift galaxies at
all. \citet{1998MNRAS.295..319M} and
\citet{1998ApJ...495..647H,2000ApJ...534..594H} resolved this issue showing
that the difficulty of detecting high-redshift DLA systems in emission is
an unavoidable consequence of the absorption cross-section selection which
tends to reveal faint galaxies because they have an integrated cross-section
larger than that of bright galaxies \citep[see also][]{1999MNRAS.305..849F}.

Recently, \citet{2004A&A...422L..33M} tentatively suggested that, if a
galaxy luminosity-metallicity relation similar to that observed
at $0\la z\la 1$
\citep[e.g.,][]{2002ApJ...581.1019G,2003ApJ...599.1006K,2004MNRAS.350..396L,2004ApJ...613..898T}
was already in place at high redshifts, then it would be possible
to significantly increase the DLA galaxy detection probability
by carefully selecting DLA systems with the highest metallicities. In fact, the
few DLA galaxies that have to date been identified in emission do give support
to the conjecture that a luminosity-metallicity relation was already in place
at $z\approx 2-3$, although the result is only marginally
statistically significant \citep[][see also Christensen et al.,
in prep.]{2004A&A...422L..33M}. This is in line with the near-Solar or even
super-Solar metallicities derived for bright Lyman-break or bright K-band
selected galaxies at similar
redshifts \citep{2004ApJ...612..108S,2004ApJ...608L..29D}. A mass-metallicity
relation has recently been put into evidence for
UV-selected star-forming galaxies at $z\sim 2.3$ by
\citet{2006astroph0602473E}.

In this paper, we provide for the first time evidence for the existence of
a velocity-metallicity correlation for high-redshift DLA galaxies that could be
the consequence of an underlying mass-metallicity relation for the
galaxies responsible for DLA absorption lines. From the observation of a sample
of 17 DLA systems at $z_{\rm abs}<3$, \citet{1998ApJ...494L..15W}
previously showed that the DLA systems exhibiting the largest line
profile velocity widths span a narrow range of high metallicities.
However, these authors also suggested that systems with small velocity
widths span a wide range of metallicities. Recently,
\citet{2003MNRAS.345..480P} found a hint of an increase of the mean DLA
metallicity with increasing velocity width, but the statistical significance of
their result is low. In this paper, we use our database of VLT-UVES quasar
spectra to build up a sample of 70 DLA or strong sub-DLA systems with total
neutral hydrogen column densities of $\log N($H\,{\sc i}$)\ga 20$ and redshifts
in the range $1.7<z_{\rm abs}<4.3$. We present new, homogeneous measurements
of DLA metallicities and line profile velocity widths in Sect.~\ref{sample} and
the observed velocity-metallicity relation in Sect.~\ref{relation}. We discuss
in Sect.~\ref{discussion} the use of the DLA gas kinematics as a proxy for
the mass of DLA galaxies and the possibility of the existence of
a mass-metallicity relation for high-$z$ DLA galaxies. We conclude in
Sect.~\ref{conclusions}.

\section{UVES DLA sample}\label{sample}

\subsection{Metallicity measurements}\label{metallicities}

Most of the systems in our sample were selected from the follow-up of the Large
Bright QSO Survey \citep[LBQS;][]{1995ApJ...454..698W} and observed at the ESO
VLT with UVES between 2000 and 2004 in the course of a systematic search for
molecular hydrogen at $z_{\rm abs}>1.8$
\citep{2000A&A...364L..26P,2003MNRAS.346..209L}. Our total sample comprises
57 DLA systems ($\log N($H\,{\sc i}$)\ge 20.3$) and 13 strong sub-DLA systems
with total neutral hydrogen column densities in the
range $20.0\la \log N($H\,{\sc i}$)<20.3$. This is only slightly lower than the
classical definition for DLA systems to ensure that these absorbers are mostly
neutral and share the same physical nature as classical DLA
systems \citep[see][]{1995MNRAS.276..268V}.

We have carefully measured or remeasured total neutral hydrogen column
densities, $\log N($H\,{\sc i}$)$, and average DLA metallicities for all the
systems in our sample. Results are summarized in Table~\ref{kinetab}.
The absorption line analysis was performed in an homogeneous manner
using standard Voigt-profile fitting techniques adopting the
oscillator strengths compiled by \citet{2003ApJS..149..205M}. For the
damping coefficients, we also adopted here the compilation
by \citet{2003ApJS..149..205M}, which results in some cases in a
slight increase of $\log N($H\,{\sc i}$)$ values compared to
\citet{2003MNRAS.346..209L}. Total metal column densities were derived as the
sum of the column densities measured in individual components of the line
profiles. Average gaseous metallicities relative to
Solar, [X/H$]\equiv\log [N($X$)/N($H$)]-\log [N($X$)/N($H$)]_\odot$,
were calculated using Solar abundances listed
in \citet{2003ApJS..149..205M}, which are based on meteoritic data from
\citet{2002ASR....30....3G}. To avoid problems related to possible
depletion onto dust grains, metallicities given in Table~\ref{kinetab} were
computed for elements that are known to deplete very little in the ISM of
the Galaxy. The reference element was taken to be X$=$Zn when Zn\,{\sc ii}
is detected, or else either S or Si was used \citep[see][for
a discussion]{2003MNRAS.346..209L}.

A noticeable property of this large dataset is that it samples well both ends
of the DLA metallicity distribution, from [X/H$]\approx -2.6$ up to about
half of Solar (see Table~\ref{kinetab}).

%------------------------------------------------------------------------------

\begin{figure*}
\centering
\includegraphics[width=8.8cm,bb=62 228 487 568,clip,angle=0]{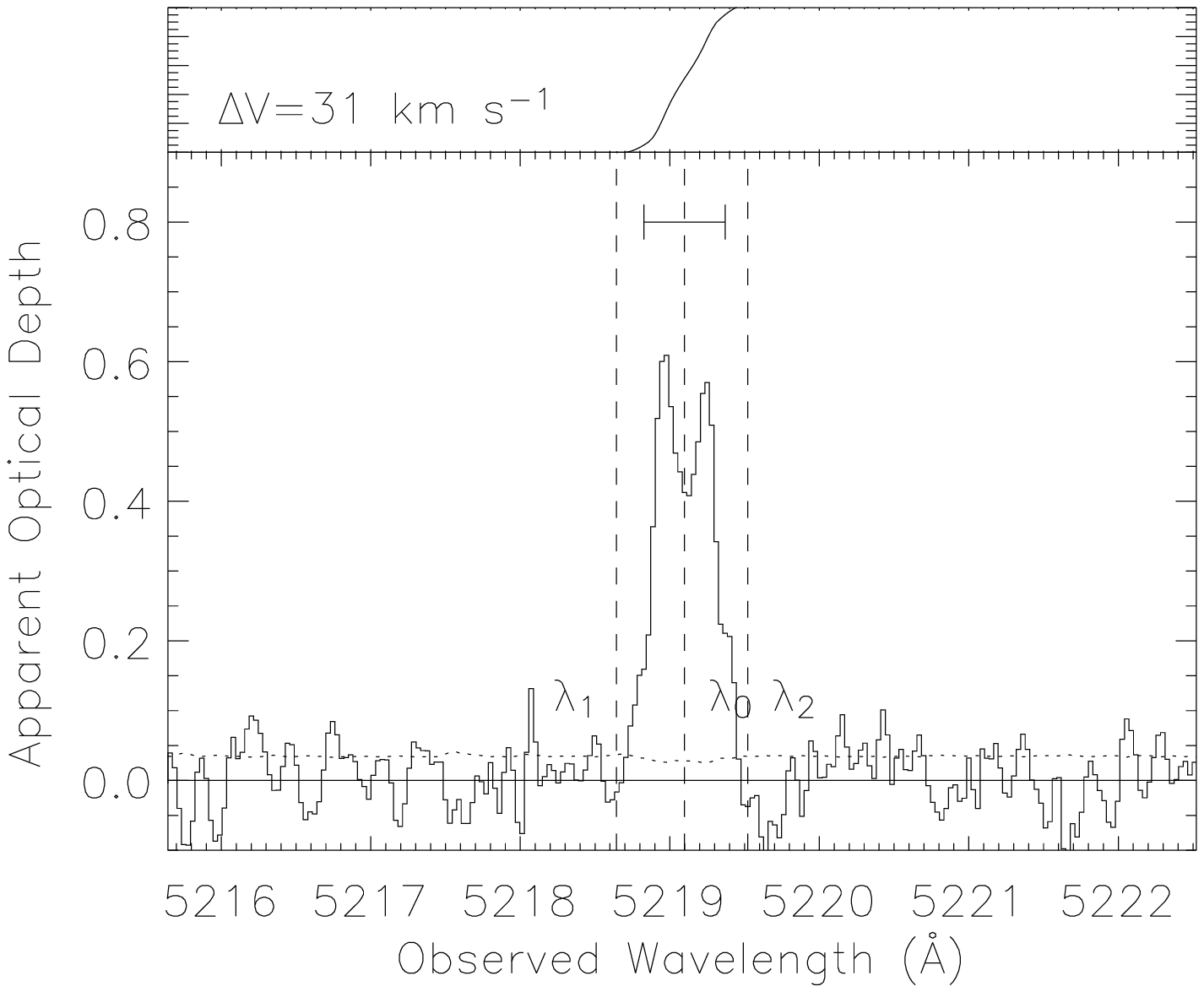}\hspace{0.25cm}\includegraphics[width=8.8cm,bb=62 228 487 568,clip,angle=0]{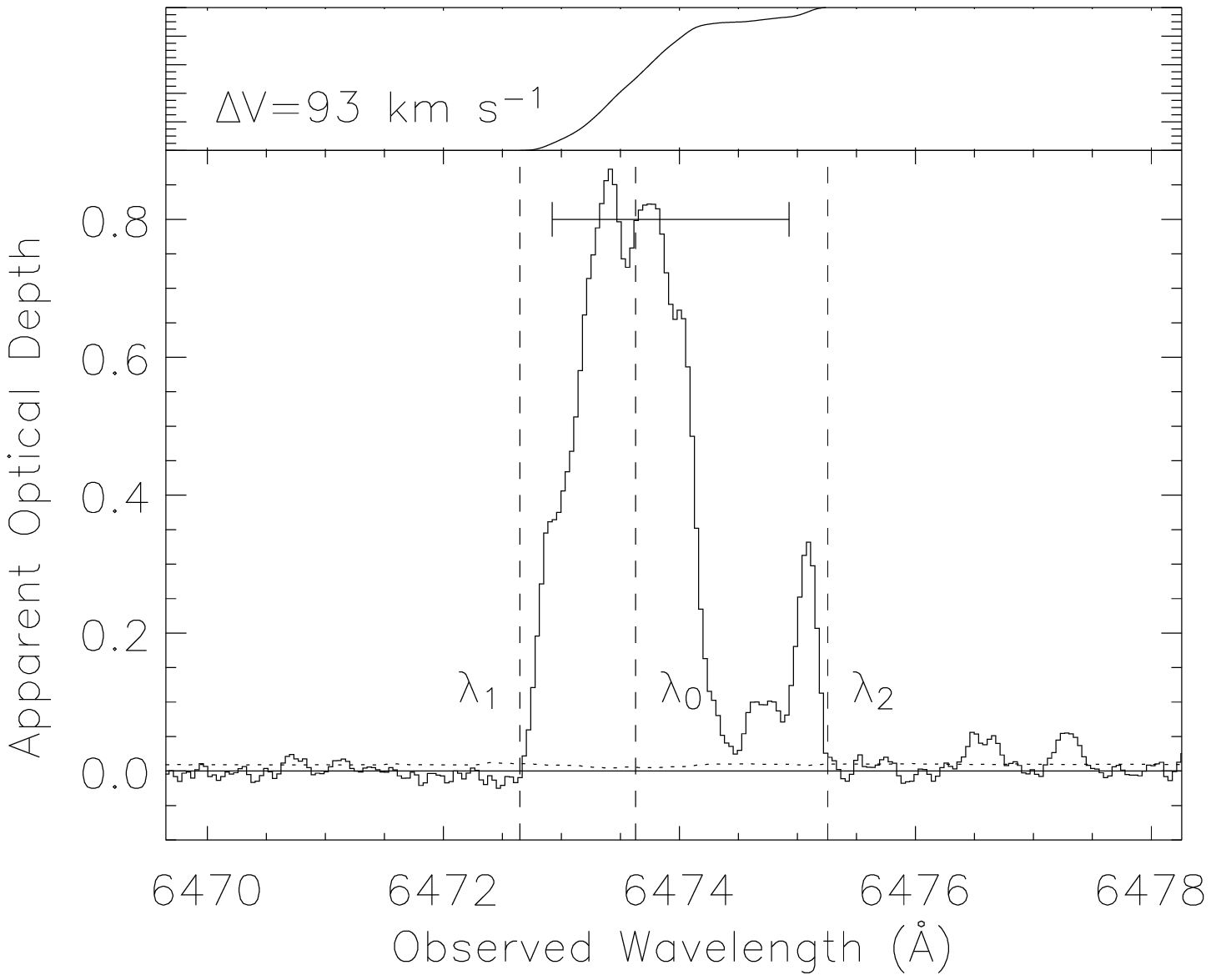}

\includegraphics[width=8.8cm,bb=62 228 487 568,clip,angle=0]{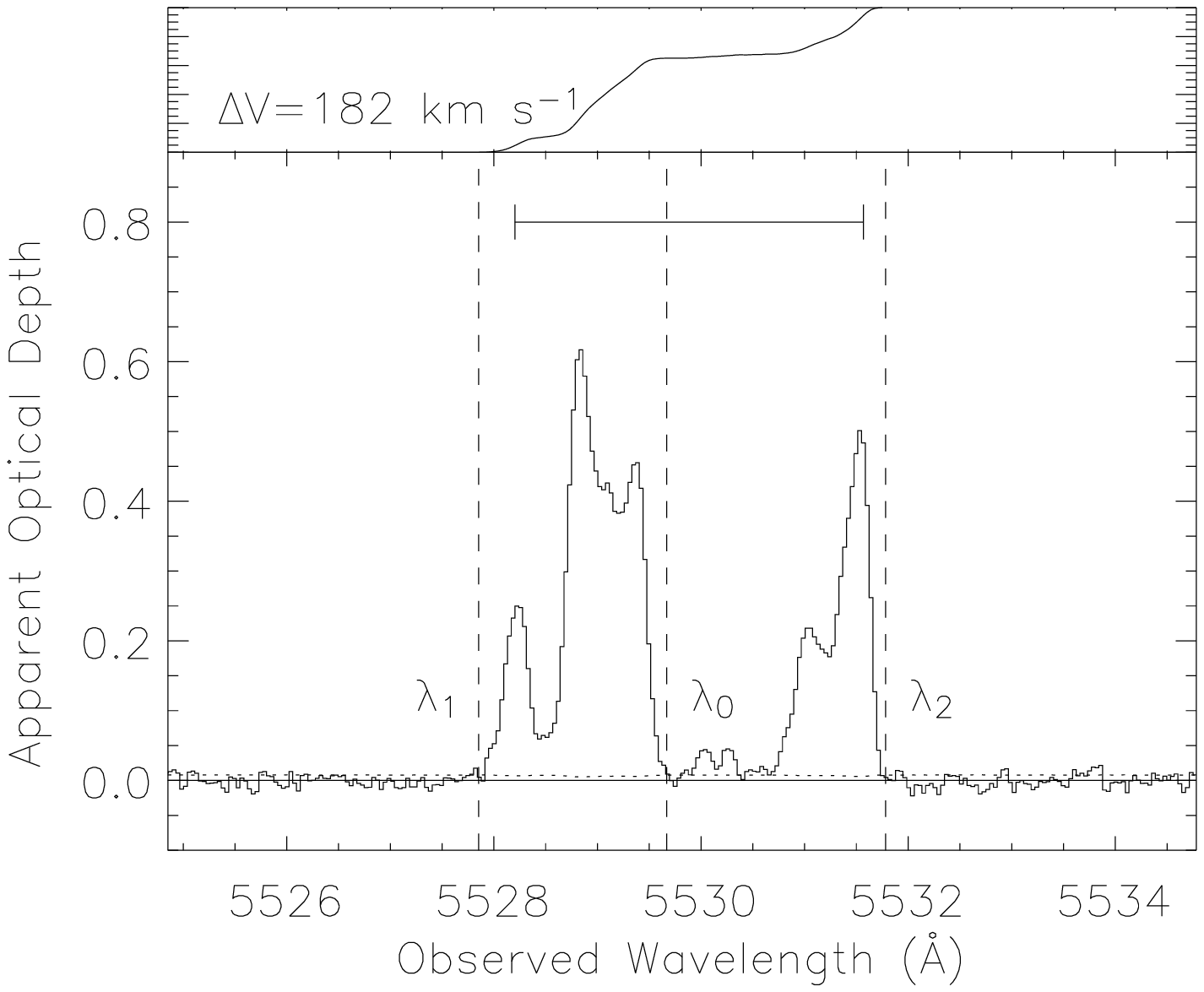}\hspace{0.25cm}\includegraphics[width=8.8cm,bb=62 228 487 568,clip,angle=0]{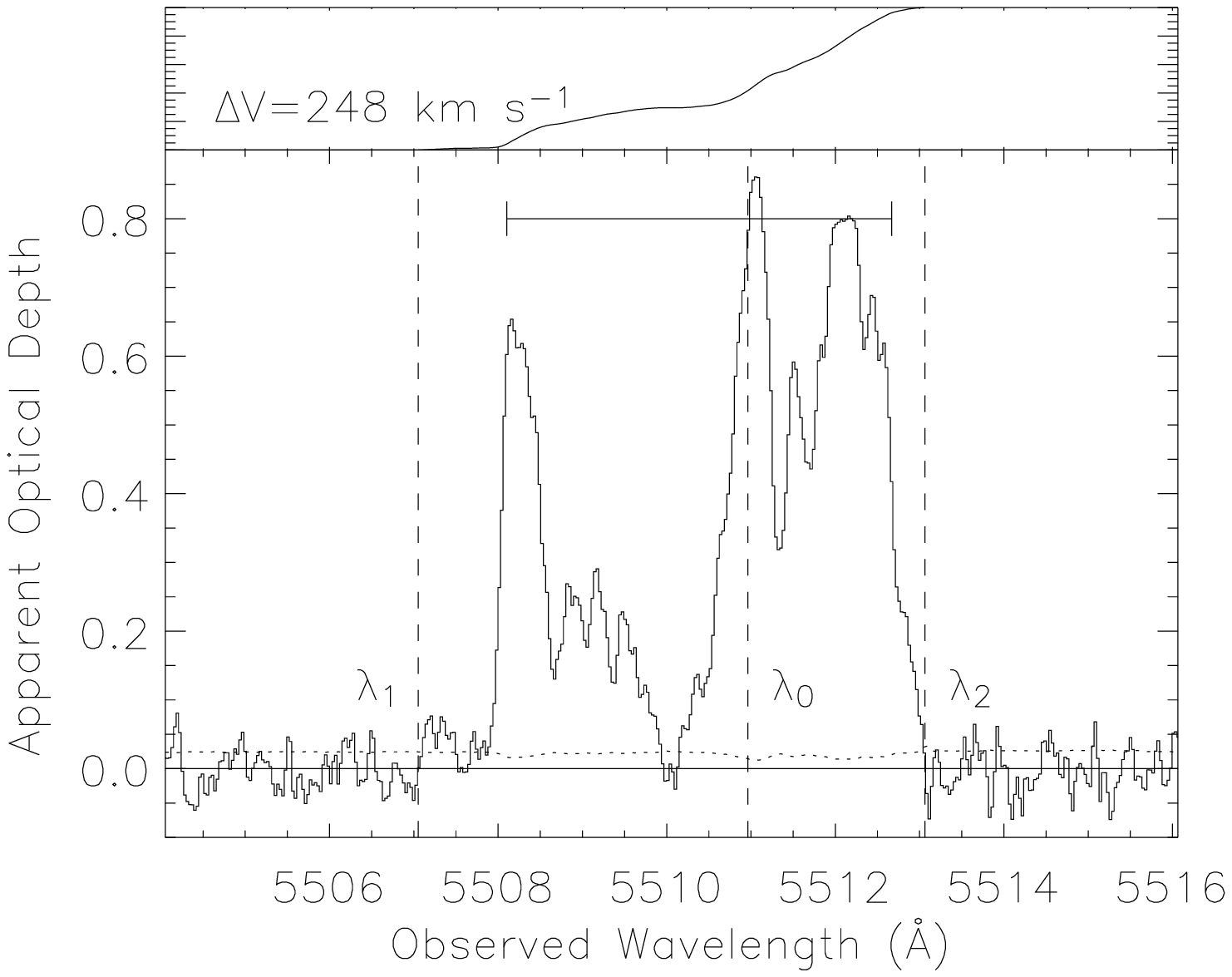}

\caption[]{
Measurements of the velocity width of low-ionization line profiles, $\Delta V$.
The different panels show the apparent optical depth in the selected transition
line for a given system: Si\,{\sc ii}$\lambda$1526 at
$z_{\rm abs}=2.418$ toward Q\,0112$-$306 (upper left),
Fe\,{\sc ii}$\lambda$1608 at $z_{\rm abs}=3.025$ toward Q\,0347$-$383 (upper
right), Si\,{\sc ii}$\lambda$1526 at $z_{\rm abs}=2.622$ toward Q\,0405$-$443
(lower left), and Fe\,{\sc ii}$\lambda$1608 at $z_{\rm abs}=2.426$ toward
Q\,2348$-$011 (lower right). The top part of each panel shows the
integration of the apparent optical depth starting at
wavelength $\lambda _1$. The horizontal bar below that corresponds to
the measurement of $\Delta V$ when 5\% of the total apparent optical depth
is avoided at both edges of the profile.}
\label{profile}
\end{figure*}

%------------------------------------------------------------------------------

\subsection{Low-ionization line kinematics}\label{kinematics}

For each of the systems, we also determined the velocity widths of
metal absorption line profiles. Low-ionization transition lines that are
not strongly saturated were selected to trace the kinematics induced
predominantly by gravity. For high-ionization lines, the velocity widths
could be dominated by peculiar ejection of hot gas. We measured the
line profile velocity widths
following \citeauthor{1997ApJ...487...73P} (\citeyear{1997ApJ...487...73P}; see
also \citealt{1998ApJ...495..647H}). We used the criterion that the residual
intensity $I$ of the strongest absorption feature in the selected line profile
must satisfy $0.1<I/I_c<0.6$, where $I_c$ is the intensity level of the
adjacent continuum. This criterion selects transitions that are
neither strongly saturated (in which case the optical depth cannot be derived
and the velocity width could be overestimated), nor too weak (in which case the
velocity width could be underestimated because part of the gas would be
undetected). For the few systems for which none of the observed lines
satisfies the above criterion, we used the mean value of the velocity
widths calculated from (i) a line slightly more saturated, and (ii) a
line slightly weaker, than what the criterion specifies
(see Table~\ref{kinetab}). From visual inspection of the
strongest low-ionization line profiles of a given system, we established the
velocity range over which the previously selected line profile should
be integrated (this corresponds to the interval [$\lambda_1$;$\lambda_2$] in
Fig.~\ref{profile}).

In order to ease comparison with previous works
\citep[e.g.,][]{1997ApJ...487...73P}, we then calculated the line profile
velocity width, $\Delta V$, as $c[\lambda(95\%)-\lambda(5\%)]/\lambda _0$,
where $\lambda(5\%)$ and $\lambda(95\%)$ are the wavelengths corresponding
to, respectively, the five per cent and 95 per cent percentiles of the apparent
optical depth distribution, and $\lambda _0$ is the first moment of this
distribution (see Fig.~\ref{profile}). Note that all our spectra have a typical
signal-to-noise ratio larger than 30. Excluding the extended wings of
the line profiles avoids taking into account satellite components that are not
strictly related to the bulk of the systems.

%------------------------------------------------------------------------------

\begin{figure}
\centering
\includegraphics[height=8.8cm,bb=45 55 569 629,clip,angle=-90]{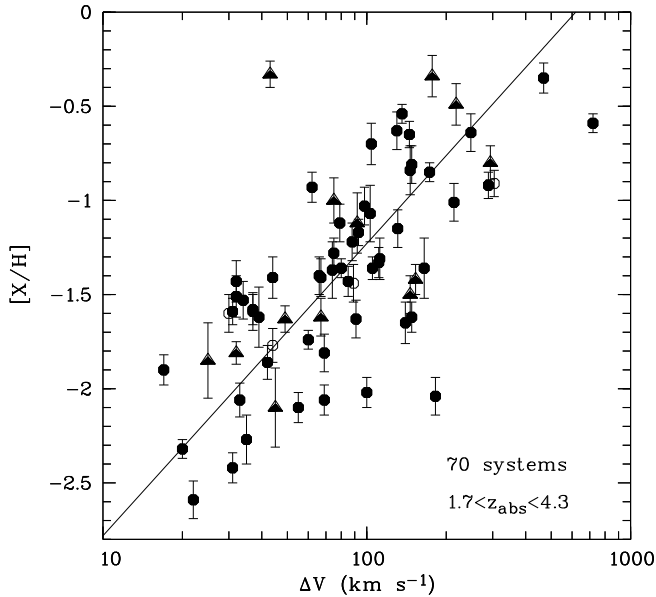}
\caption[]{
Average metallicity of each of the 70 DLA or strong sub-DLA systems in
our sample, [X/H], vs. the velocity width of their low-ionization line
profiles, $\Delta V$, displayed on a logarithmic scale. A positive
correlation between the two quantities is detected at
the $6.1\sigma$ significance level using a Kendall rank correlation test. The
linear least-square bisector fit is shown as a solid line. Sub-DLA systems,
with total neutral hydrogen column densities $20\la\log N($H\,{\sc i}$)<20.3$,
are indicated by filled triangles. There are a few DLA systems with
$c|z_{\rm em}-z_{\rm abs}|/(1+z_{\rm abs})\le 5000$ km s$^{-1}$. For the
sake of completeness, they are shown here as empty circles but they were not
considered in the analysis.}
\label{correlation}
\end{figure}

%------------------------------------------------------------------------------

\section{The velocity-metallicity relation}\label{relation}

\subsection{Detected correlation}\label{characteristics}

In Fig.~\ref{correlation}, we plot on a logarithmic scale the
average metallicity of each of the 70 DLA or strong sub-DLA systems in
our sample, [X/H], versus the velocity width of their low-ionization
line profiles, $\Delta V$. A positive correlation between the two quantities is
detected at the $6.1\sigma$ significance level using a Kendall rank correlation
test. Note that the DLA and strong sub-DLA populations are statistically
indistinguishable even though the mean metallicity is slightly larger among
sub-DLA systems. The Kolmogorov-Smirnov test probability that the two velocity
width distributions (resp. metallicity distributions) are drawn from the same
parent population is $P_{\rm KS}=0.91$ (resp. $P_{\rm KS}=0.72$) in the
two-sided case. In the following, we therefore consider the DLA and
strong sub-DLA systems in our sample altogether.

The typical measurement uncertainties in velocity width ($\pm 0.02$ dex) and
metallicity ($\pm 0.10$ dex) are small compared to the intrinsic scatter of the
data points (see Fig.~\ref{correlation}). We thus fitted the data using the
linear least-square bisector method \citep{1990ApJ...364..104I}. For
the correlation fits, we did not include the DLA systems
with $c|z_{\rm em}-z_{\rm abs}|/(1+z_{\rm abs})\le 5000$ km s$^{-1}$, where
$z_{\rm em}$ is the QSO emission redshift. However,
these $z_{\rm abs}\approx z_{\rm em}$ systems are not associated with
the central engine of the quasar, nor ejected by the quasar, but
rather associated with dense gas in its surroundings
\citep[e.g.,][]{1994A&A...291...29P,2000A&A...357..414S}. It can be seen
in Fig.~\ref{correlation} that their inclusion would not affect the
observed correlations and that their metallicities and line profile velocity
widths are representative of those of the overall DLA population.

The best-fit linear relation is:
\begin{equation}\label{eq1}
[{\rm X}/{\rm H}]=1.55(\pm 0.12)\log\Delta V -4.33(\pm 0.23)
\end{equation}
with $\Delta V$ expressed in km s$^{-1}$.

The observed high-redshift velocity-metallicity correlation presented here is
consistent with the results of \citet{2003ApJ...595L...5N} \citep[see
also][]{2005astroph0506701T}. Using the Sloan Digital Sky Survey (SDSS), these
authors found that for strong low-ionization Mg\,{\sc ii} absorbers
at $1\la z_{\rm abs}\la 2$ the Mg\,{\sc ii}$\lambda$2796 equivalent width,
or equivalently the absorption line velocity spread, is correlated with
the metallicity. In addition, they showed that, within the large
equivalent width regime, the average metallicity is larger at lower redshift.

%------------------------------------------------------------------------------

\begin{figure}
\centering
\includegraphics[height=8.8cm,bb=45 55 569 629,clip,angle=-90]{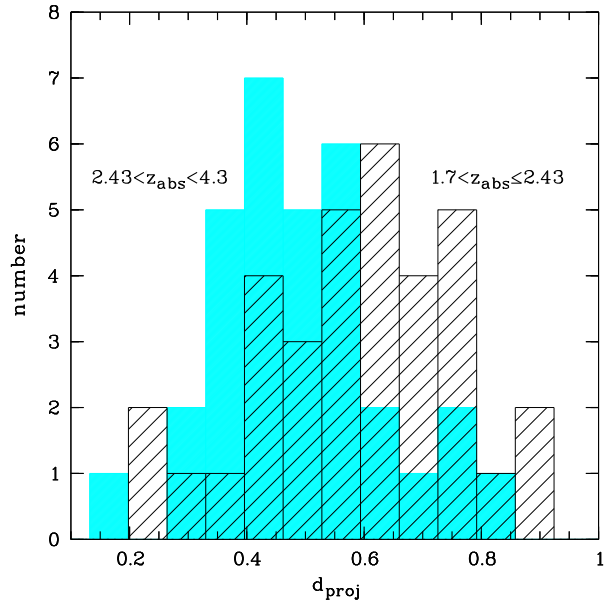}
\caption[]{
Histogram of the projected position, $d_{\rm proj}$, of each system along the
best-fit correlation relation plotted as a solid line in
Fig.~\ref{correlation}, with $d_{\rm proj}=0$ (resp. $d_{\rm proj}=1$) at the
intersection of this line with the horizontal line corresponding to
[X/H$]=-3$ (resp. [X/H$]=0$). The shaded (resp. hashed) histogram
represents the higher (resp. lower) redshift half of the sample.}
\label{projection}
\end{figure}

%------------------------------------------------------------------------------

\begin{figure*}
\centering
\includegraphics[height=8.8cm,bb=45 55 569 629,clip,angle=-90]{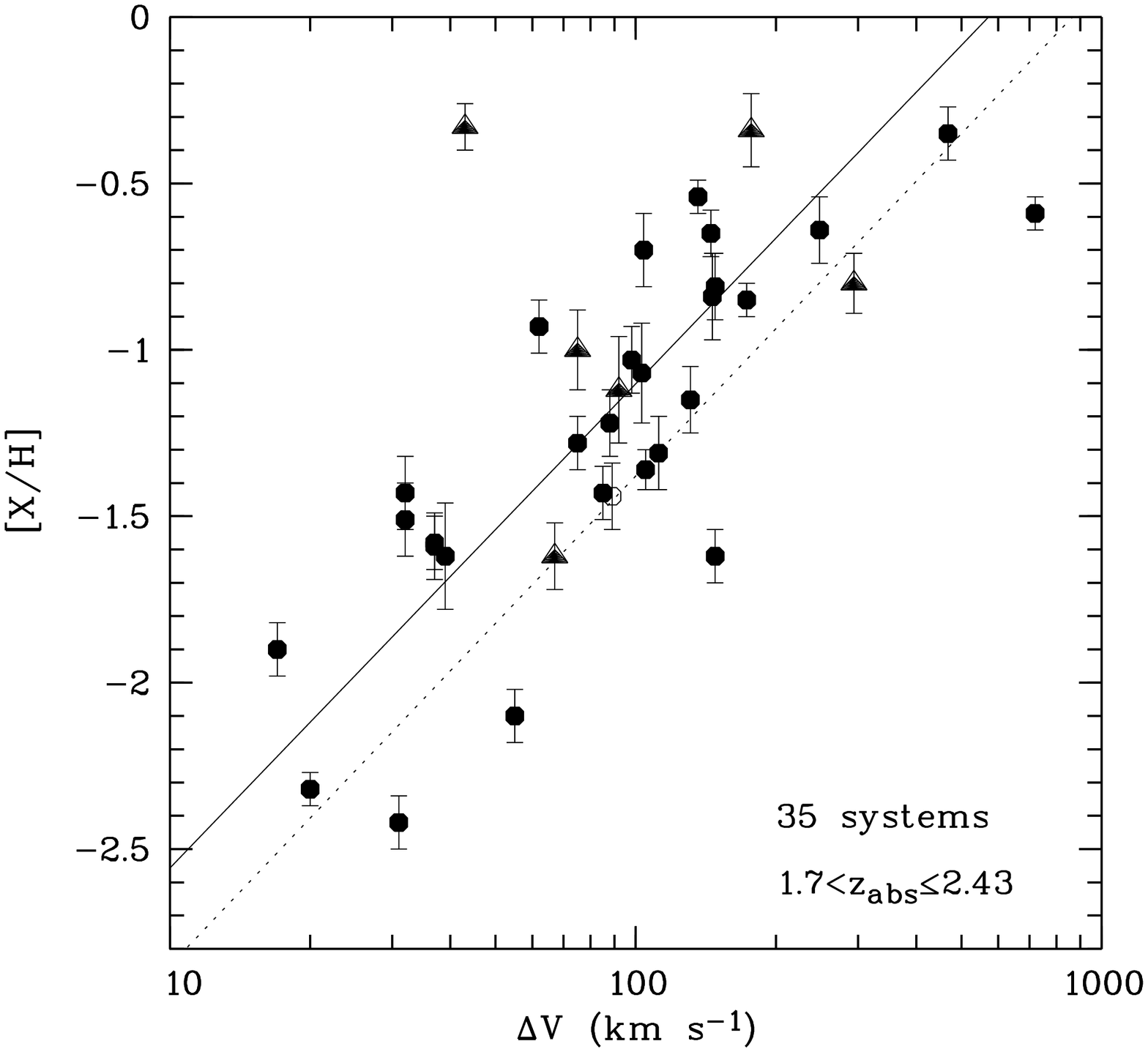}\hspace{0.25cm}\includegraphics[height=8.8cm,bb=45 55 569 629,clip,angle=-90]{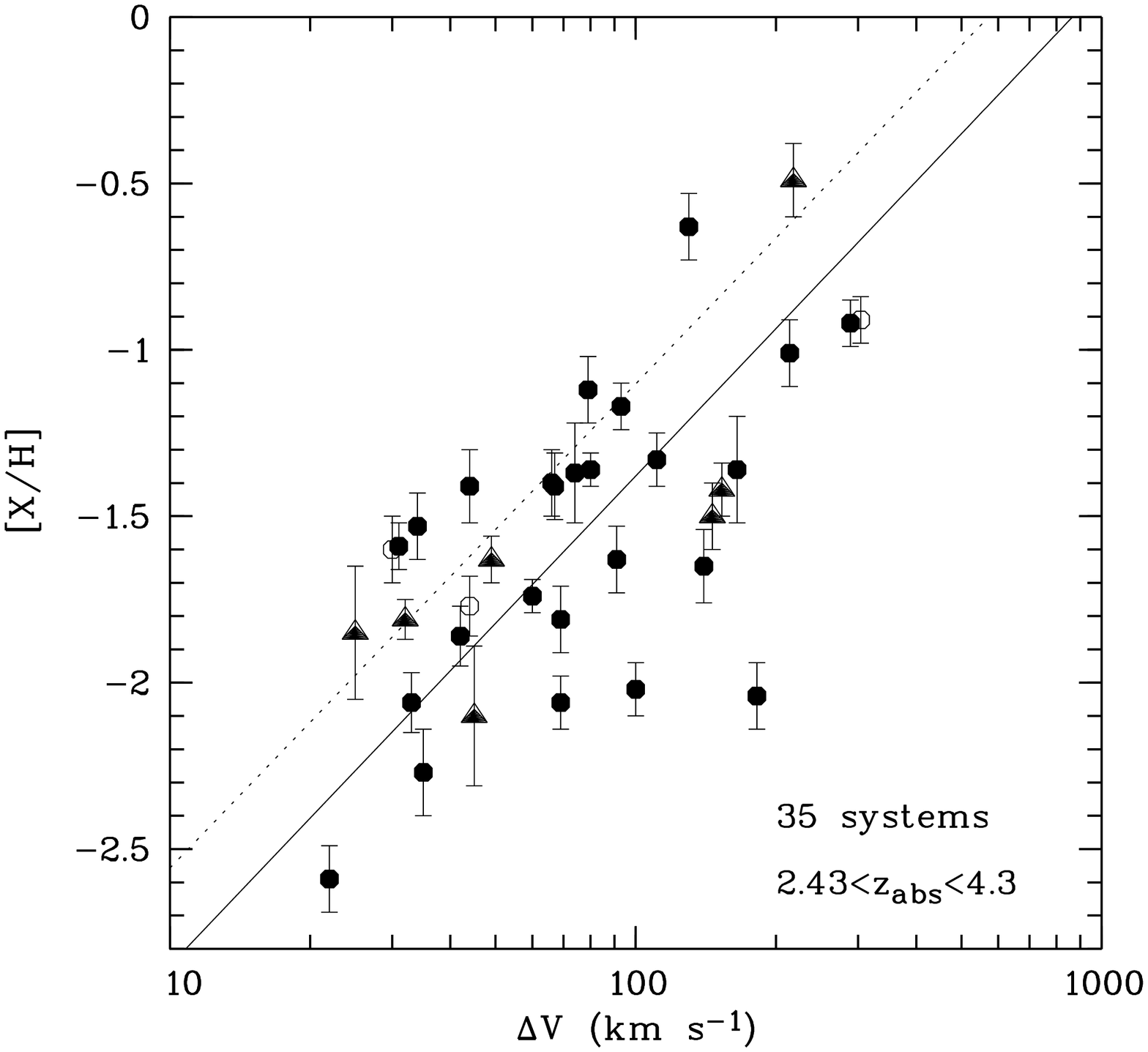}
\caption[]{
Same as Fig.~\ref{correlation} but for the sub-sample of systems with
$1.7<z_{\rm abs}\le 2.43$ ({\it left hand-side panel}) or the sub-sample
of systems with $2.43<z_{\rm abs}<4.3$ ({\it right hand-side panel}). Positive
correlations between the two quantities are detected in both the lower and
the higher redshift sub-samples with significance levels of $4.6\sigma$ and
$3.9\sigma$, respectively. Linear least-square bisector fits to each sub-sample
are shown as solid lines in the corresponding panels; the dotted lines are the
fits to the other sub-sample. There are a few DLA systems
with $c|z_{\rm em}-z_{\rm abs}|/(1+z_{\rm abs})\le 5000$ km s$^{-1}$ in
our sample. For the sake of completeness, they are shown here as empty circles
but they were not considered in the analysis.}
\label{z-evolution}
\end{figure*}

%------------------------------------------------------------------------------

\subsection{Redshift evolution}\label{zevol}

The median redshift of our sample is $z_{\rm med}=2.43$. It can be shown
that the two sub-samples of systems with, respectively, $z_{\rm abs}>2.43$
and $z_{\rm abs}\le 2.43$ differ significantly. The median DLA metallicity
and median DLA velocity width increase with decreasing redshift: [X/H$]=-1.59$
and $\Delta V=69$ km s$^{-1}$, respectively, [X/H$]=-1.15$ and $\Delta V=92$ km
s$^{-1}$, in the higher, respectively, the lower redshift half of the
sample. This resembles the point by \citet{1998ApJ...494L..15W} that the
kinematics and metallicities of the $z_{\rm abs}>3$ and $z_{\rm abs}<3$
DLA samples could show significant differences.

To investigate this behaviour further, we have calculated for each system in
our sample the projected position, $d_{\rm proj}$, along
the best-fit correlation relation ([X/H] vs. $\Delta V$) derived
in Subsect.~\ref{characteristics} and drawn as a solid line in
Fig.~\ref{correlation}. We plot in Fig.~\ref{projection} the histograms of
$d_{\rm proj}$ for the two redshift sub-samples. It is apparent that the two
histograms differ in the sense that $d_{\rm proj}$ is larger at lower
redshift. This is expected as $d_{\rm proj}$ increases with increasing
velocity width {\sl and} increasing metallicity. A Kolmogorov-Smirnov
test confirms this: the two populations have only 1\% chance to be drawn
from the same parent population. This difference is also apparent
in Fig.~\ref{z-evolution} where we plot [X/H] versus $\Delta V$ for the two
redshift sub-samples separately.

Considering the two redshift sub-samples separately, the best-fit
linear relations are:
\begin{equation}\label{eq2}
[{\rm X}/{\rm H}]=1.45(\pm 0.17)\log\Delta V -4.01(\pm 0.33)
\end{equation}
for $1.7<z_{\rm abs}\le 2.43$,
and:
\begin{equation}\label{eq3}
[{\rm X}/{\rm H}]=1.47(\pm 0.17)\log\Delta V -4.32(\pm 0.32)
\end{equation}
for $2.43<z_{\rm abs}<4.3$, with $\Delta V$ expressed in km s$^{-1}$.

The significance levels of the correlations in the lower, respectively,
the higher redshift half of the sample are $4.6\sigma$
and $3.9\sigma$, respectively (see Fig.~\ref{z-evolution}). In addition, the
Pearson correlation coefficients are $r=0.72$ and 0.63, respectively, showing
that even in the high-redshift sub-sample a linear relation is a fairly
good description of the data. It is very interesting to note that
the correlation relations do not change significantly with redshift although
there is a statistically significant increase in both [X/H]
{\sl and} $\Delta V$ with decreasing redshift. This is discussed in
Sect.~\ref{discussion}.

\subsection{Observational scatter}\label{scatter}

Scatter in the data points is expected due to random impact parameters
through the absorbing galaxies and, indeed, the scatter observed in
Fig.~\ref{correlation} is much larger than the metallicity
measurement uncertainties. Negative radial gradients in metallicity like
those observed in the discs of large nearby spirals could also contribute
to the scatter of the data points. This effect is probably not very important
however. For instance, \citet{2005ApJ...620..703C} derived from a sample of
six $z<0.65$ galaxy-DLA pairs a metallicity gradient of
only $-0.041\pm 0.012$ dex per kpc from galactic centre to 30 $h^{-1}$ kpc
radius. In addition, the magnitude of such gradients in the discs of
nearby spirals has been questioned recently \citep{2004ApJ...615..228B}.

We note that there are in our sample a few systems departing
from the general trend. The system at $z_{\rm abs}=1.973$
toward Q\,0013$-$004 has low-ionization lines extending over up to $\sim$1100
km s$^{-1}$ in velocity space \citep{2002MNRAS.332..383P}, which is much larger
than for the other systems in our sample. The DLA system at
$z_{\rm abs}=2.622$ toward Q\,0405$-$443 has a low metallicity,
[Si/H$]=-2.04$, and a comparatively large velocity width for this metallicity,
$\Delta V=182$ km s$^{-1}$ (see Fig.~\ref{profile}, lower left-hand
side panel). An inspection of these two cases indicates that
the low-ionization line profiles are clumpy, being made of four and
two well-separated clumps, respectively. These systems could arise in
galaxy groups, tidal streaming in galaxy mergers or forming galactic
structures \citep[see,
e.g.,][]{1998A&A...337...51L,1998MNRAS.301..168N,1998ApJ...495..647H,2001MNRAS.326.1475M}.
The third peculiar system, the sub-DLA system at $z_{\rm abs}=2.287$
toward Q\,2332$-$094, is the highest metallicity absorber in our sample with
[Zn/H$]=-0.33$. It has a surprisingly small velocity width for such a high
metallicity however: its profile is a blend of two sharp metal lines
resulting in a velocity width of $\Delta V=43$ km s$^{-1}$. It is
not unexpected to find such deviant cases in a sample that large.

\section{Discussion}\label{discussion}

\subsection{Kinematics as a proxy for the mass of DLA galaxies}

The existence of a DLA velocity-metallicity correlation, over more than a
factor of 100 spread in metallicity, can be understood as the consequence of an
underlying mass-metallicity relation for the galaxies responsible for DLA
absorption lines. Peculiar ejection of hot gas should indeed primarily
affect the kinematics of high-ionization lines such as C\,{\sc iv} and
Si\,{\sc iv}. Our measurements of profile velocity widths in DLA systems are
based on low-ionization lines which should instead be dominated by motions
on galactic scale governed or induced by gravity.

For disc galaxies, the rotation velocity is a direct measure of the
galaxy's total mass. Lines-of-sight that do not pass through the centre of
the galaxy will not trace the full depth of the potential well and, therefore,
will tend to show smaller velocity dispersions than the rotation velocity (see
the models by \citealt{1997ApJ...487...73P} and \citealt{1998ApJ...494L..15W}).
Hence, for random lines-of-sight through a large sample of disc galaxies, there
should be a mean relationship between mass and profile velocity width, albeit
with a large scatter induced by the range of impact parameters and
inclination angles probed by the observations.

Infall/outflow of gas, or merging of galaxy sub-clumps, will also
produce kinematically broadened line profiles with velocity widths scaling as
the infall/outflow velocities, which again scale as the depth of the
combined potential well of the galaxies or mergers. Simulations have shown that
in that case the line profile velocity width is a good indicator of the
circular velocity of the underlying dark-matter halo
with $\Delta V\sim 0.6v_{\rm c}$
\citep{1998MNRAS.301..168N,1998ApJ...495..647H,2001MNRAS.326.1475M}.
The scatter in this relation is about a factor of two and
corresponds approximately to the width of the correlation shown in
Fig.~\ref{correlation}.

In addition, there is a positive correlation between the projected stellar mass
density and the neutral hydrogen column density of DLA systems and a
good correspondence in the spatial distribution of stars and DLA systems in
simulations including star formation, supernova feedback and feedback
by galactic winds \citep[see][]{2004MNRAS.348..435N}.

\subsection{Implications and prospects}

For reasons discussed above, we assume here that the dynamical mass of
DLA galaxies is the dominant factor setting up the observed
DLA velocity-metallicity correlation.

\subsubsection{Mass-metallicity relation}

The slope of the DLA velocity-metallicity relation is the same within
uncertainties for the two redshift halves of our sample (see
Subsect.~\ref{zevol}). There is a possible increase of the intercept point of
this relation with decreasing redshift (see Fig.~\ref{z-evolution}) but this
result is not statistically significant due to the large scatter of the
data points around the mean relations (see Eqs.~\ref{eq2} and
\ref{eq3}). However, the two redshift sub-samples differ in the sense that
the median DLA metallicity {\sl and} median DLA velocity width increase
with decreasing redshift (see Subsect.~\ref{zevol}
and Fig.~\ref{projection}). This suggests that galaxy halos of a given mass
(resp. a given metallicity) are becoming more metal-rich (resp. less massive)
with time.

This is consistent with the results of \citet{2005ApJ...635..260S} who
proposed a redshift-dependent galaxy mass-metallicity relation from the
study of $0.4<z<1.0$ galaxies selected from the Gemini Deep Deep Survey and
the Canada-France Redshift Survey. Note also that a mass-metallicity relation
has recently been put into evidence at $z\sim 2.3$ for
UV-selected star-forming galaxies \citep{2006astroph0602473E}.

%Although somewhat speculative, we can tentatively convert velocity
%width measurements into mass estimates using the fact that the
%measured rest-frame equivalent width ($W_{\rm r}$) of
%the Fe\,{\sc ii}$\lambda\lambda$2586,2600 complex in galaxies correlates with
%their stellar mass as derived from $VIK$ photometry
%\citep{2005ApJ...635..260S}, $\log W_{\rm r}=0.50\log M_\star-4.44$
%with $W_{\rm r}$ expressed in \AA\ and $M_\star$ in M$_\odot$ (S. Savaglio
%2005, private communication). We can write $\Delta V\approx W_{\rm r}c/2600$,
%from which, using Eq.~\ref{eq1}, we get:
%\begin{equation}\label{eq4}
%%[{\rm X}/{\rm H}]=0.72\log M_\star -7.46
%[{\rm X}/{\rm H}]=0.77\log M_\star -8.02
%\end{equation}
%This implies that the typical mass of high-redshift DLA galaxies is of
%the order of $10^9M_\odot$, consistent with the findings by
%\citet{2005ApJ...635..260S}.

\subsubsection{Luminosity-metallicity relation}

From Cold Dark Matter (CDM) simulations, \citet{1998ApJ...495..647H} have shown
that the velocity width of DLA systems, $\Delta V$, can be related
statistically to the circular velocity of the underlying
dark-matter halo, $v_{\rm c}=(GM/r)^{1/2}$, where $M$ is the mass in a sphere
overdense by a factor of 200 compared to the mean cosmic density. They found
$\Delta V \sim 0.6v_{\rm c}$. According
to \citet{1998MNRAS.300..817H,2000ApJ...534..594H}, the luminosity function
of $z\sim 3$ galaxies can be reproduced if a simple linear scaling of the
luminosity with the mass of the dark-matter halo is
assumed, i.e., $m_R=-7.5\log (v_{\rm c}/200\ {\rm km\ s}^{-1})+m_R^0$,
where $m_R$ is the galaxy apparent $R$-band magnitude and $m_R^0=26.6$ for
the $\Lambda$-CDM model. Using our best-fit to the velocity-metallicity
relation for $1.7<z_{\rm abs}\le 2.43$ DLA systems (Eq.~\ref{eq2}),
and $z_{\rm med}=2.09$ for this sub-sample, we derive:
\begin{equation}\label{eq5}
[{\rm X}/{\rm H}]=-0.19(\pm 0.02)M_R -4.76(\pm 0.42)+0.19(\pm 0.02)K_R
\end{equation}
where $M_R$ is the galaxy absolute $R$-band magnitude and $K_R$ is
the $K$-correction in the $R$-band. Using our best-fit to
the velocity-metallicity relation for $2.43<z_{\rm abs}<4.3$ DLA
systems (Eq.~\ref{eq3}) leads to a similar result.

It is striking to note that the slope of this
DLA luminosity-metallicity relation is consistent with that derived by
\citet{2004ApJ...613..898T} for the luminosity-metallicity relation for
$z\sim 0.1$ galaxies selected from the SDSS,
$[{\rm O}/{\rm H}]=-0.185(\pm 0.001)M_B-3.452(\pm 0.018)$. The correction
from the $R$-band at high redshift to the $B$-band at low redshift
corresponds to an additional factor that corresponds to
a non-positive $K$-correction \citep{1996ApJ...467...38K} in Eq.~\ref{eq5}.
Therefore, the intercept points of the two luminosity-metallicity relations
are different in the sense that galaxies of a given luminosity (resp. a given
metallicity) are becoming more metal-rich (resp. fainter) with time.

\section{Conclusions}\label{conclusions}

Using a sample of 70 DLA or strong sub-DLA systems with total neutral
hydrogen column densities of $\log N($H\,{\sc i}$)\ga 20$ and redshifts in
the range $1.7<z_{\rm abs}<4.3$, we have shown that there is a correlation
between metallicity ([X/H]) and line profile velocity width ($\Delta V$) at
the $6.1\sigma$ significance level. The best-fit linear relation
is $[{\rm X}/{\rm H}]=1.55(\pm 0.12)\log\Delta V -4.33(\pm 0.23)$
with $\Delta V$ expressed in km s$^{-1}$. We argued that the existence of a DLA
velocity-metallicity correlation, over more than a factor of 100 spread in
metallicity, is probably the consequence of an underlying mass-metallicity
relation for the galaxies responsible for DLA absorption lines. Assuming
a simple linear scaling of the galaxy luminosity with the mass of
the dark-matter halo, we found that the slope of the DLA velocity-metallicity
relation is consistent with that of the luminosity-metallicity relation
derived for local galaxies. If the galaxy dynamical mass is indeed the
dominant factor setting up the observed DLA velocity-metallicity correlation,
then the DLA systems exhibiting the lowest metallicities among the
DLA population should, on average, be associated with galaxies of lower masses.

Eq.~\ref{eq5} implies that the more than two orders of magnitude spread in DLA
metallicity could reflect a more than ten magnitudes spread in DLA galaxy
luminosity. Even though low-mass galaxies, i.e., gas-rich dwarf galaxies,
can undergo periods of intense star formation activity and, in this case,
have high luminosities in the UV, it is a fact that, on average, they show
lower star-formation rates than more massive galaxies
(\citealt{2004MNRAS.351.1151B}; see also \citealt{2004ApJ...603...12O}).
This may well explain the difficulty of detecting high-redshift DLA galaxies
in emission \citep[e.g.,][]{2000ApJ...536...36K,2001ApJ...551...37K}.
Furthermore, the non-detection of Ly$\alpha$ emission from the
galaxies responsible for low-metallicity DLA systems, down to Ly$\alpha$ fluxes
fainter than most of the Ly$\alpha$ emitters from the deep survey
of \citet{2003A&A...407..147F}, could be a consequence of their low masses and
their correspondingly, on average, low star-formation activity.

A significant fraction of DLA galaxies could be actively forming stars for
some period of time, but due to their small masses the bursts
of star-formation would not be powerful enough and/or would have too short
life-times. Ly$\alpha$ emission would simply be too faint or would be strong
for a too short spell of time to be detected by the current generation
of 8-10\,m class telescopes. Conversely, the existence of a DLA
mass-metallicity relation can explain the recent, tentative result by
\citet{2004A&A...422L..33M} that the few DLA systems with
detected Ly$\alpha$ emission have higher than average metallicities (see
also Christensen et al., in prep.). This should be confirmed by additional deep
imaging of the fields of QSOs with carefully selected DLA absorbers.

The DLA velocity-metallicity correlation relation studied in this paper
also needs to be investigated in the context of new high-resolution
smoothed-particle hydrodynamics simulations including the effects of
feedback in a self-consistent manner \citep[see, e.g.,][]{2004MNRAS.348..435N}.

%The observed range in DLA metallicities should mainly reflect differences in
%the host galaxy's mass.
%1) possible reasons:
%-- low mass galaxies are more gas rich
%or possibly younger than more massive systems in the sense that they have not
%yet significantly enriched their ISM with metals ??
%-- low mass galaxies have higher metal losses

%------------------------------------------------------------------------------

\begin{acknowledgements}

We thank Sandra Savaglio for sharing results prior to publication. JPUF
is supported by the Danish Natural Science Research Council (SNF). PPJ and RS
gratefully acknowledge support from the Indo-French Centre for the Promotion of
Advanced Research (Centre Franco-Indien pour la Promotion de la
Recherche Avanc\'ee) under contract No. 3004-3. PPJ thanks ESO Vitacura for
hospitality during the time part of this work has been completed.

\end{acknowledgements}

%------------------------------------------------------------------------------

{\small

\bibliographystyle{bibtex/aa}
\bibliography{paper15}

}

%------------------------------------------------------------------------------

\clearpage
\begin{longtable}{llccccccc}
\caption{UVES DLA sample: average metallicities and velocity widths of low-ionisation line profiles $^\ast$}\\
\hline
\hline
Quasar & Other name & $z_{\rm em}$ & $z_{\rm abs}$ $^1$ & $\log N($H\,{\sc i}$)$ & [X/H] $^2$ &    & $\Delta V$    & Selected             \\
       &            &              &                    &                        &            & X  & (km s$^{-1}$) & transition lines $^3$\\
\hline
\endfirsthead
\caption{continued}\\
\hline
\hline
Quasar & Other name & $z_{\rm em}$ & $z_{\rm abs}$ $^1$ & $\log N($H\,{\sc i}$)$ & [X/H] $^2$ &    & $\Delta V$    & Selected             \\
       &            &              &                    &                        &            & X  & (km s$^{-1}$) & transition lines $^3$\\
\hline
\endhead
\hline
\endfoot
Q\,0000$-$263 & {\small LBQS\,0000$-$2619}    & 4.11 & 3.390 & $21.40\pm 0.08$ & $-2.06\pm 0.09$ & Zn &  33     & O\,{\sc i}$\lambda$950   \\
Q\,0010$-$002 & {\small LBQS\,0010$-$0012}    & 2.15 & 2.025 & $20.95\pm 0.10$ & $-1.43\pm 0.11$ & Zn &  32     & Si\,{\sc ii}$\lambda$1808\\
Q\,0013$-$004 & {\small LBQS\,0013$-$0029}    & 2.09 & 1.973 & $20.83\pm 0.05$ & $-0.59\pm 0.05$ & Zn & 720     & Fe\,{\sc ii}$\lambda$1608\\
Q\,0058$-$292 & {\small LBQS\,0058$-$2914}    & 3.09 & 2.671 & $21.10\pm 0.10$ & $-1.53\pm 0.10$ & Zn &  34     & Si\,{\sc ii}$\lambda$1808\\
Q\,0100$+$130 & {\small LBQS\,0100$+$1300}    & 2.69 & 2.309 & $21.35\pm 0.08$ & $-1.58\pm 0.08$ & Zn &  37     & Cr\,{\sc ii}$\lambda$2056\\
Q\,0102$-$190 & {\small LBQS\,0102$-$1902}    & 3.04 & 2.370 & $21.00\pm 0.08$ & $-1.90\pm 0.08$ & S  &  17     & Fe\,{\sc ii}$\lambda$2374\\
Q\,0102$-$190 & {\small LBQS\,0102$-$1902}    & 3.04 & 2.926 & $20.00\pm 0.10$ & $-1.50\pm 0.10$ & Si & 146     & Si\,{\sc ii}$\lambda$1526\\
Q\,0112$-$306 & {\small LBQS\,0112$-$3041}    & 2.99 & 2.418 & $20.50\pm 0.08$ & $-2.42\pm 0.08$ & Si &  31     & Si\,{\sc ii}$\lambda$1526\\
Q\,0112$-$306 & {\small LBQS\,0112$-$3041}    & 2.99 & 2.702 & $20.30\pm 0.10$ & $-0.49\pm 0.11$ & Si & 218     & Fe\,{\sc ii}$\lambda$1608\\
Q\,0112$+$030 & {\small LBQS\,0112$+$0300}    & 2.81 & 2.423 & $20.90\pm 0.10$ & $-1.31\pm 0.11$ & S  & 112     & Fe\,{\sc ii}$\lambda$1608\\
Q\,0135$-$273 & {\small CTS\,1005}            & 3.21 & 2.107 & $20.30\pm 0.15$ & $-1.12\pm 0.16$ & S  &  82/103 & {\small Fe\,{\sc ii}$\lambda$2260/Fe\,{\sc ii}$\lambda$2586}\\
Q\,0135$-$273 & {\small CTS\,1005}            & 3.21 & 2.800 & $21.00\pm 0.10$ & $-1.40\pm 0.10$ & S  &  65/67  & {\small S\,{\sc ii}$\lambda$1259/Fe\,{\sc ii}$\lambda$1608}\\
Q\,0216$+$080 & {\small LBQS\,0216$+$0803}    & 2.99 & 1.769 & $20.30\pm 0.10$ & $-1.00\pm 0.12$ & Zn &  75     & Fe\,{\sc ii}$\lambda$2374\\
Q\,0216$+$080 & {\small LBQS\,0216$+$0803}    & 2.99 & 2.293 & $20.50\pm 0.10$ & $-0.70\pm 0.11$ & Zn & 104     & Si\,{\sc ii}$\lambda$1808\\
Q\,0336$-$017 & {\small LBQS\,0336$-$0142}    & 3.20 & 3.062 & $21.10\pm 0.10$ & $-1.41\pm 0.10$ & Si &  67     & Si\,{\sc ii}$\lambda$1808\\
Q\,0347$-$383 & {\small LBQS\,0347$-$3819}    & 3.22 & 3.025 & $20.73\pm 0.05$ & $-1.17\pm 0.07$ & Zn &  93     & Fe\,{\sc ii}$\lambda$1608\\
Q\,0405$-$443 & {\small CTS\,247}             & 3.02 & 1.913 & $20.80\pm 0.10$ & $-1.03\pm 0.10$ & Zn &  98     & Si\,{\sc ii}$\lambda$1808\\
Q\,0405$-$443 & {\small CTS\,247}             & 3.02 & 2.550 & $21.15\pm 0.15$ & $-1.36\pm 0.16$ & Zn & 165     & Si\,{\sc ii}$\lambda$1808\\
Q\,0405$-$443 & {\small CTS\,247}             & 3.02 & 2.595 & $21.05\pm 0.10$ & $-1.12\pm 0.10$ & Zn &  79     & Si\,{\sc ii}$\lambda$1808\\
Q\,0405$-$443 & {\small CTS\,247}             & 3.02 & 2.622 & $20.45\pm 0.10$ & $-2.04\pm 0.10$ & Si & 182     & Si\,{\sc ii}$\lambda$1526\\
Q\,0450$-$131 & ....                          & 2.25 & 2.067 & $20.50\pm 0.07$ & $-1.62\pm 0.08$ & S  & 148     & Fe\,{\sc ii}$\lambda$1608\\
Q\,0458$-$020 & {\small LBQS\,0458$-$0203}    & 2.29 & 2.040 & $21.70\pm 0.10$ & $-1.22\pm 0.10$ & Zn &  88     & Cr\,{\sc ii}$\lambda$2056\\
Q\,0528$-$250 & {\small LBQS\,0528$-$2505}    & 2.77 & 2.141 & $20.98\pm 0.05$ & $-1.36\pm 0.06$ & Zn & 105     & Si\,{\sc ii}$\lambda$1808\\
Q\,0528$-$250 & {\small LBQS\,0528$-$2505}    & 2.77 & 2.811 & $21.35\pm 0.07$ & $-0.91\pm 0.07$ & Zn & 304     & S\,{\sc ii}$\lambda$1253 \\
Q\,0551$-$366 & ....                          & 2.32 & 1.962 & $20.70\pm 0.08$ & $-0.35\pm 0.08$ & Zn & 468     & Si\,{\sc ii}$\lambda$1808\\
Q\,0841$+$129 & ....                          & 2.50 & 1.864 & $21.00\pm 0.10$ & $-1.51\pm 0.11$ & S  &  32     & S\,{\sc ii}$\lambda$1259 \\% makes use of better archival data
Q\,0841$+$129 & ....                          & 2.50 & 2.375 & $21.05\pm 0.10$ & $-1.59\pm 0.10$ & Zn &  37     & Fe\,{\sc ii}$\lambda$1125\\% makes use of better archival data
Q\,0841$+$129 & ....                          & 2.50 & 2.476 & $20.80\pm 0.10$ & $-1.60\pm 0.10$ & Zn &  30     & S\,{\sc ii}$\lambda$1259 \\% makes use of better archival data
Q\,0913$+$072 & {\small LBQS\,0913$+$0715}    & 2.78 & 2.618 & $20.35\pm 0.10$ & $-2.59\pm 0.10$ & Si &  22     & Si\,{\sc ii}$\lambda$1526\\
Q\,1036$-$229 & {\small CTS\,460}             & 3.13 & 2.778 & $20.93\pm 0.05$ & $-1.36\pm 0.05$ & S  &  80     & Fe\,{\sc ii}$\lambda$1081\\
%Q\,1037$-$270 & {\small Tol\,1037$-$2703}     & 2.19 & 2.139 & $19.70\pm 0.05$ & $-0.31\pm 0.05$ & Zn &  68     & Si\,{\sc ii}$\lambda$1808\\
Q\,1108$-$077 & {\small BRI\,1108$-$0747}     & 3.92 & 3.482 & $19.95\pm 0.07$ & $-1.63\pm 0.07$ & Si &  49     & Si\,{\sc ii}$\lambda$1526\\% slightly below the log N(HI)=20 threshold
Q\,1108$-$077 & {\small BRI\,1108$-$0747}     & 3.92 & 3.608 & $20.37\pm 0.07$ & $-1.59\pm 0.07$ & Si &  31     & Fe\,{\sc ii}$\lambda$1608\\
Q\,1111$-$152 & {\small CTS\,298}             & 3.37 & 3.266 & $21.30\pm 0.05$ & $-1.65\pm 0.11$ & Zn & 140     & Si\,{\sc ii}$\lambda$1020\\
Q\,1117$-$134 & {\small BR\,1117$-$1329}      & 3.96 & 3.350 & $20.95\pm 0.10$ & $-1.41\pm 0.11$ & Zn &  43/45  & {\small Si\,{\sc ii}$\lambda$1808/Fe\,{\sc ii}$\lambda$1608}\\
Q\,1157$+$014 & ....                          & 1.99 & 1.944 & $21.80\pm 0.10$ & $-1.44\pm 0.10$ & Zn &  89     & Fe\,{\sc ii}$\lambda$2260\\
Q\,1209$+$093 & {\small LBQS\,1209$+$0919}    & 3.30 & 2.584 & $21.40\pm 0.10$ & $-1.01\pm 0.10$ & Zn & 214     & Si\,{\sc ii}$\lambda$1808\\
Q\,1210$+$175 & {\small LBQS\,1210$+$1731}    & 2.54 & 1.892 & $20.70\pm 0.08$ & $-0.93\pm 0.08$ & S  &  62     & S\,{\sc ii}$\lambda$1253 \\
Q\,1223$+$178 & {\small LBQS\,1223$+$1753}    & 2.94 & 2.466 & $21.40\pm 0.10$ & $-1.63\pm 0.10$ & Zn &  91     & Si\,{\sc ii}$\lambda$1808\\
Q\,1232$+$082 & {\small LBQS\,1232$+$0815}    & 2.57 & 2.338 & $20.90\pm 0.08$ & $-1.43\pm 0.08$ & S  &  85     & Si\,{\sc ii}$\lambda$1808\\
Q\,1331$+$170 & ....                          & 2.08 & 1.776 & $21.15\pm 0.07$ & $-1.28\pm 0.08$ & Zn &  75     & Fe\,{\sc ii}$\lambda$2374\\
Q\,1337$+$113 & {\small LBQS\,1337$+$1121}    & 2.92 & 2.508 & $20.12\pm 0.05$ & $-1.81\pm 0.06$ & Si &  32     & Fe\,{\sc ii}$\lambda$2344\\
Q\,1337$+$113 & {\small LBQS\,1337$+$1121}    & 2.92 & 2.796 & $21.00\pm 0.08$ & $-1.86\pm 0.09$ & Si &  42     & Fe\,{\sc ii}$\lambda$1608\\
Q\,1340$-$136 & {\small CTS\,325}             & 3.20 & 3.118 & $20.05\pm 0.08$ & $-1.42\pm 0.08$ & S  & 153     & Fe\,{\sc ii}$\lambda$1608\\
Q\,1409$+$095 & {\small LBQS\,1409$+$0930}    & 2.85 & 2.019 & $20.65\pm 0.10$ & $-1.62\pm 0.16$ & Zn &  39     & Fe\,{\sc ii}$\lambda$2374\\
Q\,1409$+$095 & {\small LBQS\,1409$+$0930}    & 2.85 & 2.456 & $20.53\pm 0.08$ & $-2.06\pm 0.08$ & Si &  69     & Si\,{\sc ii}$\lambda$1526\\
%Q\,1409$+$095 & {\small LBQS\,1409$+$0930}    & 2.85 & 2.668 & $19.80\pm 0.08$ & $-1.41\pm 0.09$ & S  &  24     & Si\,{\sc ii}$\lambda$1304\\% nice N(OI) determination +SII...
Q\,1441$+$276 & {\small PSS J\,1443$+$2724}   & 4.42 & 4.224 & $20.95\pm 0.10$ & $-0.63\pm 0.10$ & S  & 130     & Fe\,{\sc ii}$\lambda$1081\\
Q\,1444$+$014 & {\small LBQS\,1444$+$0126}    & 2.21 & 2.087 & $20.25\pm 0.07$ & $-0.80\pm 0.09$ & Zn & 294     & Fe\,{\sc ii}$\lambda$1608\\
Q\,1451$+$123 & {\small LBQS\,1451$+$1223}    & 3.25 & 2.255 & $20.35\pm 0.10$ & $-1.07\pm 0.15$ & Zn & 103     & Fe\,{\sc ii}$\lambda$2374\\
Q\,1451$+$123 & {\small LBQS\,1451$+$1223}    & 3.25 & 2.469 & $20.40\pm 0.10$ & $-2.27\pm 0.13$ & Si &  35     & Fe\,{\sc ii}$\lambda$2382\\
Q\,1451$+$123 & {\small LBQS\,1451$+$1223}    & 3.25 & 3.171 & $20.20\pm 0.20$ & $-2.10\pm 0.21$ & Si &  45     & Si\,{\sc ii}$\lambda$1526\\
Q\,2059$-$360 & ....                          & 3.09 & 2.507 & $20.29\pm 0.07$ & $-1.85\pm 0.20$ & S  &  25     & Fe\,{\sc ii}$\lambda$2344\\
Q\,2059$-$360 & ....                          & 3.09 & 3.083 & $20.98\pm 0.08$ & $-1.77\pm 0.09$ & S  &  44     & Fe\,{\sc ii}$\lambda$1121\\
Q\,2116$-$358 & {\small HES\,2116$-$3550}     & 2.34 & 1.996 & $20.10\pm 0.07$ & $-0.34\pm 0.11$ & Zn & 177     & Fe\,{\sc ii}$\lambda$1608\\
Q\,2138$-$444 & {\small LBQS\,2138$-$4427}    & 3.17 & 2.383 & $20.60\pm 0.05$ & $-1.15\pm 0.10$ & Zn & 131     & Fe\,{\sc ii}$\lambda$2374\\
Q\,2138$-$444 & {\small LBQS\,2138$-$4427}    & 3.17 & 2.852 & $20.98\pm 0.05$ & $-1.74\pm 0.05$ & Zn &  58/62  & {\small Si\,{\sc ii}$\lambda$1808/Fe\,{\sc ii}$\lambda$2374}\\
%Q\,2152$+$137 & {\small PSS J\,2155$+$1358}   & 4.26 & 3.142 & $19.80\pm 0.10$ & $-2.10\pm 0.10$ & Si &  17     & O\,{\sc i}$\lambda$1302  \\
Q\,2152$+$137 & {\small PSS J\,2155$+$1358}   & 4.26 & 3.316 & $20.50\pm 0.15$ & $-1.37\pm 0.15$ & Si &  74     & Fe\,{\sc ii}$\lambda$1608\\
%Q\,2152$+$137 & {\small PSS J\,2155$+$1358}   & 4.26 & 4.212 & $19.70\pm 0.10$ & $-1.81\pm 0.10$ & Si &  40     & O\,{\sc i}$\lambda$1302  \\
Q\,2206$-$199 & {\small LBQS\,2206$-$1958}    & 2.56 & 1.921 & $20.67\pm 0.05$ & $-0.54\pm 0.05$ & Zn & 136     & Si\,{\sc ii}$\lambda$1808\\
Q\,2206$-$199 & {\small LBQS\,2206$-$1958}    & 2.56 & 2.076 & $20.44\pm 0.05$ & $-2.32\pm 0.05$ & Si &  20     & Si\,{\sc ii}$\lambda$1304\\
Q\,2230$+$025 & {\small LBQS\,2230$+$0232}    & 2.15 & 1.864 & $20.90\pm 0.10$ & $-0.81\pm 0.10$ & S  & 148     & S\,{\sc ii}$\lambda$1253 \\
Q\,2231$-$002 & {\small LBQS\,2231$-$0015}    & 3.02 & 2.066 & $20.55\pm 0.07$ & $-0.65\pm 0.07$ & S  & 145     & S\,{\sc ii}$\lambda$1253 \\
Q\,2243$-$605 & {\small HES\,2243$-$6031}     & 3.01 & 2.331 & $20.65\pm 0.05$ & $-0.85\pm 0.05$ & Zn & 173     & Si\,{\sc ii}$\lambda$1808\\%good  53000 resolution
Q\,2332$-$094 & {\small FIRST J\,2334$-$0908} & 3.32 & 2.287 & $20.07\pm 0.07$ & $-0.33\pm 0.07$ & Zn &  43     & Si\,{\sc ii}$\lambda$1808\\
Q\,2332$-$094 & {\small FIRST J\,2334$-$0908} & 3.32 & 3.057 & $20.50\pm 0.07$ & $-1.33\pm 0.08$ & S  & 111     & Fe\,{\sc ii}$\lambda$1608\\
%Q\,2342$+$342 & ....                          & 3.01 & 2.90  &                 &                 &    &         & ...new data to analysed...companion to the Red is a LLS?? \\
Q\,2343$+$125 & ....                          & 2.51 & 2.431 & $20.40\pm 0.07$ & $-0.92\pm 0.07$ & Zn & 289     & Fe\,{\sc ii}$\lambda$1608\\
Q\,2344$+$125 & ....                          & 2.76 & 2.538 & $20.50\pm 0.10$ & $-1.81\pm 0.10$ & Si &  69     & Fe\,{\sc ii}$\lambda$1608\\
Q\,2348$-$011 & {\small LBQS\,2348$-$0108}    & 3.01 & 2.426 & $20.50\pm 0.10$ & $-0.64\pm 0.10$ & S  & 248     & Fe\,{\sc ii}$\lambda$1608\\
Q\,2348$-$011 & {\small LBQS\,2348$-$0108}    & 3.01 & 2.615 & $21.30\pm 0.08$ & $-2.02\pm 0.08$ & Si &  95/105 & {\small Si\,{\sc ii}$\lambda$1808/Si\,{\sc ii}$\lambda$1526}\\
Q\,2348$-$147 & ....                          & 2.94 & 2.279 & $20.63\pm 0.05$ & $-2.10\pm 0.08$ & S  &  55     & Fe\,{\sc ii}$\lambda$2586\\
Q\,2359$-$022 & {\small LBQS\,2359$-$0216}    & 2.81 & 2.095 & $20.65\pm 0.10$ & $-0.84\pm 0.13$ & Zn & 146     & Fe\,{\sc ii}$\lambda$1608\\
Q\,2359$-$022 & {\small LBQS\,2359$-$0216}    & 2.81 & 2.154 & $20.30\pm 0.10$ & $-1.62\pm 0.10$ & Si &  67     & Si\,{\sc ii}$\lambda$1526
%Q\,2332$-$094 & {\small FIRST J\,2334$-$0908} & & 2.152 &                 &           & & & Si\,{\sc ii}$\lambda$1808\\ upper limit on N(HI) due to Lya to the Blue of a LLS
%J\,2344       &                               & &       &                 &           & & &                          \\ very bad spectrum
%              & UM\,681                       & & 1.78  & $19.4$          &           & & &                          \\ not bad but real sub-DLA...
%%
\label{kinetab}
\end{longtable}
\flushleft
$^\ast$ Total neutral hydrogen column density and metallicity measurements
are from this work, except for the systems at $z_{\rm abs}=3.390$
toward Q\,0000$-$263 \citep{2001ApJ...549...90M}, $z_{\rm abs}=3.025$ and 2.087
toward, respectively, Q\,0347$-$383
and Q\,1444$+$014 \citep{2003MNRAS.346..209L}, $z_{\rm abs}=1.962$
toward Q\,0551$-$366 \citep{2002A&A...392..781L}, and
$z_{\rm abs}=4.224$ toward Q\,1441$+$276 \citep{2006ApJ...640L..25L}.\\
Quoted uncertainties are $1\sigma$ standard deviations.\newline

$^1$ There are a few DLA systems
with $c|z_{\rm em}-z_{\rm abs}|/(1+z_{\rm abs})<5000$ km s$^{-1}$. These
systems were not considered during the correlation analyzes. However, their
properties (metallicity, $N($H\,{\sc i}$)$, ionisation level) do not resemble
those of the associated systems. They are not associated with the central
engine of the quasar, nor ejected by the quasar, but rather associated
with dense gas in its surroundings (see text).\\
$^2$ Average gaseous metallicities relative to Solar,
[X/H$]\equiv\log [N($X$)/N($H$)]-\log [N($X$)/N($H$)]_\odot$, calculated
using Solar abundances listed in \citet{2003ApJS..149..205M}, which are based
on meteoritic data from \citet{2002ASR....30....3G}. The reference element was
taken to be X$=$Zn when Zn\,{\sc ii} is detected, or else either S or Si
was used.\\
$^3$ Transition lines used to determine the velocity widths
of low-ionisation line profiles. Whenever none of the observed lines satisfies
the adopted criterion (see Sect.~\ref{sample}), we used the mean value of
the velocity widths calculated from (i) a line slightly more saturated, and
(ii) a line slightly weaker, than what the criterion specifies.

%------------------------------------------------------------------------------

\end{document}